\documentstyle[tighten,aps,epsf,twocolumn]{revtex}

\begin{document}
\draft

\def\bra#1{{\langle#1\vert}}
\def\ket#1{{\vert#1\rangle}}
\def\coeff#1#2{{\scriptstyle{#1\over #2}}}
\def\undertext#1{{$\underline{\hbox{#1}}$}}
\def\hcal#1{{\hbox{\cal #1}}}
\def\sst#1{{\scriptscriptstyle #1}}
\def\eexp#1{{\hbox{e}^{#1}}}
\def\rbra#1{{\langle #1 \vert\!\vert}}
\def\rket#1{{\vert\!\vert #1\rangle}}
\def\lsim{{ <\atop\sim}}
\def\gsim{{ >\atop\sim}}
\def\nubar{{\bar\nu}}
\def\psibar{{\bar\psi}}
\def\Gmu{{G_\mu}}
\def\alr{{A_\sst{LR}}}
\def\wpv{{W^\sst{PV}}}
\def\evec{{\vec e}}
\def\notq{{\not\! q}}
\def\notk{{\not\! k}}
\def\notp{{\not\! p}}
\def\notpp{{\not\! p'}}
\def\notder{{\not\! \partial}}
\def\notcder{{\not\!\! D}}
\def\notA{{\not\!\! A}}
\def\notv{{\not\!\! v}}
\def\Jem{{J_\mu^{em}}}
\def\Jana{{J_{\mu 5}^{anapole}}}
\def\nue{{\nu_e}}
\def\mn{{m_\sst{N}}}
\def\mns{{m^2_\sst{N}}}
\def\me{{m_e}}
\def\mes{{m^2_e}}
\def\mq{{m_q}}
\def\mqs{{m_q^2}}
\def\mz{{M_\sst{Z}}}
\def\mzs{{M^2_\sst{Z}}}
\def\ubar{{\bar u}}
\def\dbar{{\bar d}}
\def\sbar{{\bar s}}
\def\qbar{{\bar q}}
\def\sstw{{\sin^2\theta_\sst{W}}}
\def\gv{{g_\sst{V}}}
\def\ga{{g_\sst{A}}}
\def\pv{{\vec p}}
\def\pvs{{{\vec p}^{\>2}}}
\def\ppv{{{\vec p}^{\>\prime}}}
\def\ppvs{{{\vec p}^{\>\prime\>2}}}
\def\qv{{\vec q}}
\def\qvs{{{\vec q}^{\>2}}}
\def\xv{{\vec x}}
\def\xpv{{{\vec x}^{\>\prime}}}
\def\yv{{\vec y}}
\def\tauv{{\vec\tau}}
\def\sigv{{\vec\sigma}}
\def\sst#1{{\scriptscriptstyle #1}}
\def\gpnn{{g_{\sst{NN}\pi}}}
\def\grnn{{g_{\sst{NN}\rho}}}
\def\gnnm{{g_\sst{NNM}}}
\def\hnnm{{h_\sst{NNM}}}

\def\xivz{{\xi_\sst{V}^{(0)}}}
\def\xivt{{\xi_\sst{V}^{(3)}}}
\def\xive{{\xi_\sst{V}^{(8)}}}
\def\xiaz{{\xi_\sst{A}^{(0)}}}
\def\xiat{{\xi_\sst{A}^{(3)}}}
\def\xiae{{\xi_\sst{A}^{(8)}}}
\def\xivtez{{\xi_\sst{V}^{T=0}}}
\def\xivteo{{\xi_\sst{V}^{T=1}}}
\def\xiatez{{\xi_\sst{A}^{T=0}}}
\def\xiateo{{\xi_\sst{A}^{T=1}}}
\def\xiva{{\xi_\sst{V,A}}}

\def\rvz{{R_\sst{V}^{(0)}}}
\def\rvt{{R_\sst{V}^{(3)}}}
\def\rve{{R_\sst{V}^{(8)}}}
\def\raz{{R_\sst{A}^{(0)}}}
\def\rat{{R_\sst{A}^{(3)}}}
\def\rae{{R_\sst{A}^{(8)}}}
\def\rvtez{{R_\sst{V}^{T=0}}}
\def\rvteo{{R_\sst{V}^{T=1}}}
\def\ratez{{R_\sst{A}^{T=0}}}
\def\rateo{{R_\sst{A}^{T=1}}}

\def\mro{{m_\rho}}
\def\mks{{m_\sst{K}^2}}
\def\mpi{{m_\pi}}
\def\mpis{{m_\pi^2}}
\def\mom{{m_\omega}}
\def\mphi{{m_\phi}}
\def\Qhat{{\hat Q}}

\def\FOS{{F_1^{(s)}}}
\def\FTS{{F_2^{(s)}}}
\def\GAS{{G_\sst{A}^{(s)}}}
\def\GES{{G_\sst{E}^{(s)}}}
\def\GMS{{G_\sst{M}^{(s)}}}
\def\GATEZ{{G_\sst{A}^{\sst{T}=0}}}
\def\GATEO{{G_\sst{A}^{\sst{T}=1}}}
\def\mdax{{M_\sst{A}}}
\def\mustr{{\mu_s}}
\def\rsstr{{r^2_s}}
\def\rhostr{{\rho_s}}
\def\GEG{{G_\sst{E}^\gamma}}
\def\GEZ{{G_\sst{E}^\sst{Z}}}
\def\GMG{{G_\sst{M}^\gamma}}
\def\GMZ{{G_\sst{M}^\sst{Z}}}
\def\GEn{{G_\sst{E}^n}}
\def\GEp{{G_\sst{E}^p}}
\def\GMn{{G_\sst{M}^n}}
\def\GMp{{G_\sst{M}^p}}
\def\GAp{{G_\sst{A}^p}}
\def\GAn{{G_\sst{A}^n}}
\def\GA{{G_\sst{A}}}
\def\GETEZ{{G_\sst{E}^{\sst{T}=0}}}
\def\GEIEZ{{G_\sst{E}^{\sst{I=0}}}}
\def\GETEO{{G_\sst{E}^{\sst{T}=1}}}
\def\GMTEZ{{G_\sst{M}^{\sst{T}=0}}}
\def\GMTEO{{G_\sst{M}^{\sst{T}=1}}}
\def\lamd{{\lambda_\sst{D}^\sst{V}}}
\def\lamn{{\lambda_n}}
\def\lams{{\lambda_\sst{E}^{(s)}}}
\def\bvz{{\beta_\sst{V}^0}}
\def\bvo{{\beta_\sst{V}^1}}
\def\Gdip{{G_\sst{D}^\sst{V}}}
\def\GdipA{{G_\sst{D}^\sst{A}}}
\def\fks{{F_\sst{K}^{(s)}}}
\def\FIS{{F_i^{(s)}}}
\def\fpi{{F_\pi}}
\def\fk{{F_\sst{K}}}

\def\RAp{{R_\sst{A}^p}}
\def\RAn{{R_\sst{A}^n}}
\def\RVp{{R_\sst{V}^p}}
\def\RVn{{R_\sst{V}^n}}
\def\rva{{R_\sst{V,A}}}
\def\xbb{{x_B}}

\def\etal{{{\em et al.}}}

\def\delalr{{{delta\alr\over\alr}}}
\def\pbar{{\bar{p}}}
\def\lamchi{{\Lambda_\chi}}
\def\rsstr{{\langle r_s^2\rangle}}
\def\bpp{{b_1^{1/2,\ 1/2}}}
\def\bpm{{b_1^{1/2,\ -1/2}}}
\def\bll{{b_1^{\lambda, \lambda'}}}
\def\fks{{F_K^{(s)}}}

\def\simge{\hspace*{0.2em}\raisebox{0.5ex}{$>$}
     \hspace{-0.8em}\raisebox{-0.3em}{$\sim$}\hspace*{0.2em}}
\def\simle{\hspace*{0.2em}\raisebox{0.5ex}{$<$}
     \hspace{-0.8em}\raisebox{-0.3em}{$\sim$}\hspace*{0.2em}}

\title{
\hfill{\normalsize INT \#DOE/ER/40561-323-INT97-00-170}\\
\hfill{\normalsize MKPH-T-97-14}\\[2.5cm]
$KN$ Scattering and the Nucleon Strangeness Radius}

\author{M.J. Ramsey-Musolf$^{a,}$\thanks{On leave from the
Department of Physics, University of Connecticut, Storrs, CT
08629\ \ USA}
and H.-W. Hammer$^{a,b,}$\thanks{Present address: TRIUMF, 4004 Wesbrook Mall, 
Vancouver, B.C., Canada V6T 2A3}\\[0.3cm]
}
\address{
$^a$ Institute for Nuclear Theory, 
University of Washington, Seattle, WA 98195 USA\\
$^b$Universit{\"a}t Mainz, Institut f{\"u}r Kernphysik,
D-55099 Mainz, Germany
}


\maketitle

\begin{abstract}
The leading non-zero electric moment of the nucleon strange-quark
vector current is the mean square strangeness radius, $\rsstr$. We evaluate 
the lightest OZI-allowed contribution to $\rsstr$, arising from the
kaon cloud, using dispersion relations. Drawing upon unitarity constraints
as well as $K^{+}N$ scattering and $e^+e^-\to K\bar{K}$ cross
section data, we find the structure of this contribution differs significantly
from that suggested by a variety of QCD-inspired model calculations. In
particular, we find evidence for a strong $\phi$-meson resonance which may
enhance the scale of kaon cloud contribution to an observable level.
\end{abstract}


\narrowtext

The reasons for the success of the non-relativistic quark model (NRQM)
in accounting for the static properties of low-lying hadrons remains
one of the un-solved mysteries of non-perturbative QCD. It has been
postulated that -- as far as low-energy observables are concerned --
constituent quarks of the NRQM effectively account for the QCD degrees of 
freedom inside hadrons\cite{Kap88}. This suggestion may be tested, in part, by
measuring observables which depend only on the sea quarks and gluons,
such as the matrix element $\bra{N}\bar{s}\gamma_\mu s\ket{N}$. By now
there exists a well-defined program of parity-violating (PV) electron-nucleus
scattering experiments dedicated to the determination of this matrix 
element\cite{Mus94}.
The theoretical understanding of this matrix element is much
less clear. To date, only two
results from the lattice have been reported\cite{Lei96}, 
and they appear to conflict with the
recent results for the strangeness magnetic form factor reported by the
SAMPLE collaboration\cite{Mue97}. 
The use of hadronic effective theory, in the guise
of chiral perturbation theory, is also limited, as chiral symmetry
does not afford an independent determination of the relevant low-energy
constants\cite{Ram97}. 
A third alternative -- the use of QCD-inspired models -- is
equally problematic, and model predictions for the mean square strangeness
radius ($\rsstr$) and magnetic moment ($\mustr$) vary considerably in
magnitude and sign (for a recent review of model calculations, 
see Ref.~\cite{Ram97}). 

        In this Letter, we analyze the strangeness radius using a fourth 
approach, namely, dispersion relations (DR's). Our objective is to 
identify the hadronic mechanisms which govern the leading strangeness 
moments without relying on QCD-inspired nucleon models. To that end, we focus
on the lightest OZI-allowed contribution, which arises from the so-called
\lq\lq kaon cloud" or $K\bar{K}$ intermediate state. A variety of
model calculations
reported to date have assumed that (a) OZI-allowed processes, in the guise of
virtual strange intermediate states, give the most important contributions
to the leading strangeness moments; and (b) these processes are adequately
described by truncating at second order in the strong hadronic couplings, $g$.
In the case of $\rsstr$, the resultant predictions are generally smaller
than would be observable in the parity-violating (PV) electron scattering
experiments.  Moreover, kaon cloud predictions are typically an order of
magnitude smaller than those obtained with DR's under the assumption of
of vector meson dominance (VMD), which relies
on the extraction of a large $\phi NN$ coupling from analyses of the
isoscalar EM form factors\cite{Jaf89}.
Assumption (a) has been analyzed elsewhere and shown to
be questionable \cite{Ha97}. Regarding assumption (b),
we find that the structure of the strangeness
form factors differs significantly from the assumptions underlying kaon cloud
models, particularly the validity of the second-order approximation. 
Using the kaon cloud as an illustrative case study, we show that the
scale of OZI-allowed contributions depends critically on effects going
beyond ${\cal O}(g^2)$, and that
a proper inclusion of such effects may enhance the
scale of OZI-allowed contributions to an experimentally detectable level.
We also demonstrate the relationship between resonance \cite{Jaf89} and
kaon cloud contributions, resolving a long-standing issue in this field.

Of the form factors which parameterize $\bra{N}\bar{s}\gamma_\mu s\ket{N}
$, we focus on the strangeness electric form factor $\GES$, for which
we obtain our most reliable results. Since $\GES$ vanishes at $q^2=0$, we 
write a subtracted dispersion relation for this form factor and its leading, 
non-vanishing moment:
\begin{eqnarray}
\label{eq:gesdr}
\GES(t)&=&{t\over\pi}\int_{t_0}^\infty\ dt'\ {{\hbox{Im}}\ \GES(t')\over
        t'(t'-t)}\ \ \ ,\\
\label{eq:rsstrdr}
\rsstr = \left.6{d\GES\over dt}\right\vert_{t=0}&=&{6\over\pi}
\int_{t_0}^\infty\ dt'\ {{\hbox{Im}}\ \GES(t')\over t^{\prime 2}}\ \ \ ,
\end{eqnarray}
where $t=q^2$ and 
$t_0$ begins a cut along the real $t$-axis associated with a
given physical intermediate state. The state having the lowest threshold
$t_0$ is the $3\pi$ state, whose contribution is nominally OZI-violating.
The lightest OZI-allowed contribution arises from the $K\bar{K}$ state,
for which $t_0=4\mks$. Following the analysis of Ref. \cite{Mu97}, we express
the $K\bar{K}$ contribution to the absorptive part of the electric form
factor as a product of the appropriate $K\bar{K}\to N\bar{N}$ partial
wave and the kaon strangeness vector current form factor:
\begin{equation}
\label{eq:gessf}
{\hbox{Im}}\ \GES(t) = {\hbox{Re}}\ \left\{\left({Q\over
4\mn}\right)\bpp(t)\fks(t)^{\ast}\right\}\ \ \ .
\end{equation}
Here, $Q=\sqrt{t/4-\mks}$, $\fks$ parameterizes the matrix element
$\bra{0}\bar{s}\gamma_\mu s\ket{K(k_1)\bar{K}(k_2)}$, and $\bll$
is the $J=1$ partial wave for $K\bar{K}$ to scatter to 
the state $\ket{N(\lambda)\bar{N}(\lambda')}$ with $\lambda, \lambda'$
denoting the corresponding helicities. 

The problem now is to determine $\bpp$ and $\fks$ as reliably as possible.
For $t\geq 4\mns$, the physical $N\bar{N}$ production threshold, one may
in principle use $K\bar{K}\to N\bar{N}$ data to determine
the scattering amplitude. Alternatively, we note that in this kinematic
region, the unitarity of the $S$-matrix implies that $|\bll|\leq 1$\cite{Mu97}.
Given the present quality of $K\bar{K}\to N\bar{N}$ scattering data,
it turns out to be more effective to insert the unitarity bound on
$\bpp$ into Eq. (\ref{eq:gessf}). The corresponding bound on the 
contribution from $t\geq 4\mns$ to the dispersion integral of 
Eq. (\ref{eq:rsstrdr}) is negligible (see Table \ref{tab1}). 

To evaluate $\bpp$ in the unphysical region $4\mks\leq t\leq 4\mns$, one
must rely on some method of analytic continuation. The ${\cal O}(g^2)$
(one-loop) model
calculations of $\GES$ implicity rely on an analytic expression to effect
this continuation. The one kaon-loop approximation is equivalent to using a DR
in which the $\bll$ are computed in the Born Approximation (BA) and the kaon 
strangeness form factor taken to be point-like [$\fks(t)\equiv -1$]\cite{Mu97}. 
The BA calcuation yields an analytic expression for $\bpp$, which may be
analytically continued to unphysical values of $t$. 
When computed in this approximation, however, the $\bll$ violate the 
unitarity bound by a factor of four or more for $t\geq 4\mns$ \cite{Mu97}. 
This violation reflects
the omission of important kaon rescattering and other short-distance hadronic
mechanisms which render the scattering amplitude consistent with the 
requirements of unitarity. Given the magnitude of the unitarity violation
at the physical threshold, one would infer that the BA represents
a rather drastic approximation in the unphysical region and that truncation
at ${\cal O}(g^2)$ is questionable.

An alternate strategy, which we follow here, is to perform a fit to experimental
$K^+ N$ scattering amplitudes and analytically continue the results 
for the fit into the unphysical region. 
The success of this approach depends on the quality of the data,
the kinematic range over which it exists, and the
stability of the fit. It is advantageous to consider 
$K^+ N$ (s-channel) amplitudes, since the analytic continuation may 
be performed without encountering problematic singularities occuring 
in the the u-channel reaction $K^- N$. As experimental input, we use 
the recent phase shift analysis of the VPI-group \cite{Hy92}.
The requisite $K^+ N$ amplitudes 
of sufficient quality exists over the range $-8\mks\simle t < 0$. We
correspondingly expect our continuation to yield a credible estimate
of the scattering amplitude for $0\leq t\simle 8\mks$. Although this
range does not include the entire unphysical region over which one
must compute the dispersion integral, it is sufficiently broad for 
our present purposes. Indeed, the structure we find in the continued
amplitude in this region appreciably affects the kaon cloud contribution
to $\rsstr$.

We carry out the continuation by using backward dispersion
relations. This method relies on the coincidence
of s- and t-channel amplitudes in the kinematic domain:
$\cos\theta_{\rm s}=\cos\theta_{\rm t}=-1$ for a given 
value of $t$,
where $\theta_{\rm s}$ and $\theta_{\rm t}$ are the c.m. 
scattering
angles. Thus, we may continue the s-channel amplitude to the unphysical 
region in $t$, and equate this amplitude with the corresponding t-channel
amplitude: $A(\cos\theta_{\rm s}=-1, t_{\rm phys})\rightarrow
A(\cos\theta_{\rm s}=-1, t_{\rm unphys})=
A(\cos\theta_{\rm t}=-1, 
t_{\rm unphys})$. In performing the continuation of the backward 
s-channel amplitude, we follow conventional procedures \cite{Ni70,Ni72}
and work with the discrepancy function, $\Delta A$:
\begin{eqnarray}
\label{eq:discrep}
\Delta A(t)&=&A(t) -A^{\rm pole}(t)
-{{\cal P}\over\pi}\int^{t_p}_0\ dt' {{\hbox{Im}} 
A^{\rm exp}(t')\over t'-t} \\ &-&{1\over\pi} \int^{-\infty}_{t_p} 
\ dt'{{\hbox{Im}} A^{\rm model}(t')\over t'-t}\ \ \ , \nonumber
\end{eqnarray}
where data exist in the range $t_p\leq t\leq 0$, $A^{\rm exp}(t)$ gives 
the experimental amplitude in this range, $A^{\rm model}(t)$ gives
a model for the high-energy part of the amplitude, and
$A^{\rm pole}(t)$ denotes the
$\Lambda$, $\Sigma$ pole terms, which are known exactly 
(the dependence on $\cos\theta_{\rm s}$ is implicit).  The function 
$\Delta A$ is analytic for real values of $t$ in the range $t_p\leq t\leq 
0$, whereas the full amplitude has a cut in this region.
For $A^{\rm model}$, we use a Regge exchange model fit \cite{Ba69}. 
The inferred high-$t$ behavior is $A^{\rm model}(t) \propto |t|^\alpha$, 
with $\alpha=-1.2$, and the constant factor
is determined from the highest-energy experimental point.

The continuation proceeds by replacing $A(t)\to{\hbox{Re}}\ 
A^{\rm exp}(t)$
in Eq.~(\ref{eq:discrep}) and 
performing a fit to the experimental values
for $\Delta A$ in the range $t_p\leq t\leq 0$. This fit is most easily
accomplished by first carrying out a conformal map $t\to z(t)$, transforming
the domain of analyticity for $\Delta A$ into a circle of unit radius 
\cite{Vi75}. A convergent expansion of the amplitude is then made using 
Legendre polynomials, $P_k (z)$, and the coefficients determined from a fit 
to the experimental values for $\Delta A$. In practice, we fit 
$g(z(t))=\Delta A/|t|^\alpha$, which divides out the high energy behavior
of $\Delta A$ and improves the stability of the continuation. 
The discrepancy function -- and, via  Eq. 
(\ref{eq:discrep}), the full backward
amplitude -- is then obtained in the unphysical 
region  by evaluating the Legendre exansion for $z=z(t_{\rm unphys})$. 

The final step in obtaining $\bpp$ is to expand 
$A(\cos\theta_{t}=-1, t_{\rm unphys})$ in partial waves, 
using the helicity basis as in Ref. \cite{Mu97,Ni72}. 
Since the amplitude is known only for one value of $\theta_{t}$, 
a separation of the $\bll$ from the $b_{J\not= 1}^{\lambda,\lambda'}$
relies on several additional observations:
(a) the only significant enhancements of the $J\geq 2$ partial waves
occur via resonances;
(b) the lightest $J>1$, $I=0$ resonance having a non-negligible branching
ratio to the $K\bar{K}$ state is the $f_2(1270)$, whose mass lies near
the upper end of the range in $\sqrt{t}$ for which we expect our continuation
to be valid;  and (c) the use of a realistic parameterization for $\fks$ 
in Eq. (\ref{eq:gessf}) dramatically suppresses contributions for $\sqrt{t}
\simge 1.4 {\hbox{ GeV}}/c$. We thus expect little contamination from 
the $f_2(1270)$ and higher-lying resonances
in the region near the $K\bar{K}$ threshold, which gives the dominant
contribution to $\rsstr$, and we correspondingly omit the $J\geq 2$ partial
waves.

The remaining $J=0$ and $J=1$ partial waves
may be separated by drawing on the work of Refs. \cite{Ni72,Br79}. 
In those analyses, the strength and approximate peak position 
of the $J=0$ partial wave were determined from $K N$ phase 
shift analyses using backward dispersion relations and
its generalization. 

As a check on our general procedure, we reproduce
the results of Ref. \cite{Ni70} for the $\pi N$ case.
Furthermore, we have tested the sensitivity of our results for
$\rsstr$ to changes of the high-energy parameter $\alpha$ and
the truncation point of the fitted Legendre series, $n$. We find
that the structure of the continued amplitudes and resultant value
of $\rsstr$ change by $\simle 15\%$ 
as $\alpha$ and $n$ are varied over reasonable
ranges. We obtain the results shown in Fig. 1 and Table I using
$n=6$ and the value for $\alpha$ taken from the Regge model fit \cite{Ba69}, 
$\alpha=-1.2$.

\begin{figure}

\vspace{-3.1cm}

\begin{center}~
\epsfxsize=9cm
\epsffile{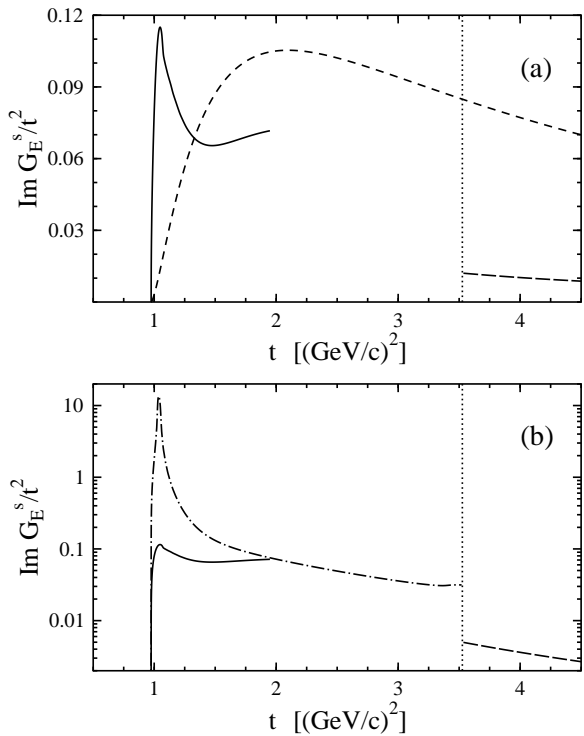}
\end{center}

\vspace{-3.5cm}

\caption{\label{fig1} $K\bar{K}$ contribution to the spectral function for
$\rsstr$ in units of $({\hbox{GeV}}/c)^{-4}$. 
Short-dashed curve (a)  gives results for ${\cal O}(g^2)$ calculation . Solid
curve [(a) and (b)] gives AC/PFF results, while long-dashed curves
show unitarity bound for $t\geq 4\mns$ using PFF (a) and GS 
form factor (b). 
Dash-dotted curve gives
AC/GS spectral function (b). Dotted vertical line indicates physical
$N\bar{N}$ production threshold.}
\end{figure}

The results of this analysis are displayed in Fig. \ref{fig1}a, 
where we plot the kaon cloud contribution to the
spectral function for $\rsstr$ as a function of $t$. We give
the spectral function computed using a point-like form factor (PFF) for
$\fks$ and two scenarios for $\bpp$: (I) the BA, and (II) analytic
continuation (AC) of $K^+ N$ scattering amplitudes. We also show the upper 
limit on the spectral function generated by the unitarity bound on $\bpp$
for $t\geq 4\mns$. The curve obtained in scenario (II) contains
a peak in the vicinity of the $\phi(1020)$
meson, presumably reflecting the presence of a $K\bar{K}\leftrightarrow
\phi$ resonance in the scattering amplitude.
This structure enhances the spectral function over the result
obtained at ${\cal O}(g^2)$ near the beginning of the $K\bar{K}$ cut. 
As $t$ increases from
$4\mks$, the spectral function obtained in scenario (II) falls below that
of scenario (I), ostensibly due to $K\bar{K}$ rescattering which must
eventually bring the spectral function below the unitarity bound for
$t\geq 4\mns$. As observed previously in Ref. \cite{Mu97}, 
and as illustrated in Fig. \ref{fig1}a, the ${\cal O}(g^2)$ approximation 
omits these rescattering corrections and consequently
violates the unitarity bound by a factor of four or more even at threshold.

Turning to the kaon strangeness form factor, we note that the assumption
of point-like behavior [$\fks(t)\equiv -1$] is poorly justified on
phenomenological grounds. Data for $e^+e^{-}\to K\bar{K}$ indicate that
the kaon EM form factor is strongly peaked for $t\approx m_\phi^2$
and falls off sharply from unity for $t\simge 2$ $({\hbox{GeV}}/c)^2$. Although
there exist no data for $\fks(t)$, one expects it to behave in an analogous 
fashion to $F_K^{EM}(t)$ \cite{Mu97}. We therefore follow Ref. \cite{Mu97} 
and employ a Gounaris-Sakurai (GS) parameterization for $\fks$, which 
produces the correct normalization at $t=0$ and a peak in the vicinity of 
the $\phi$ meson. The corresponding spectral function is shown in 
Fig. \ref{fig1}b,
where comparison is made with the spectral function computed using a 
point-like $\fks$. The use of the GS parameterization significantly
enhances the spectral function near the beginning of the $K\bar{K}$ cut
as compared with the point-like case, while it suppresses the spectral
function for $t\simge 2\ ({\hbox{GeV}}/c)^2$.
We note that other parameterizations
for $\fks$, such as a simple $\phi$-dominance form, yield similar results
for the spectral function.

The numerical consequences of using experimental $K^+ N$ amplitudes to 
determine $\bpp$
and of employing a realistic kaon strangeness form factor are indicated in
Table \ref{tab1}. The first line gives the ${\cal O}(g^2)$
kaon cloud prediction for 
$\rsstr$, using a point-like $\fks$ and $\bpp$ computed in the BA.
The second line gives results when the GS form factor and 
analytically continued fit to $K^+ N$ amplitudes for $\bpp$ are used 
(AC/GS). In both cases the unitarity bound on $\bpp$ is imposed for
$t\geq 4\mns$. The AC/GS results were obtained by extending $\bpp$ to
$4\mns$. Although we believe the continuation to be trustworthy only for
$t\simle 8\mks$, and although the continued $\bpp$ amplitude exceeds the
unitarity bound for $t\to 4\mns$,
the GS form factor suppresses contributions
for $t\geq 8\mks$, rendering the overall contribution from $8\mks\leq t\leq
4\mns$ negligible. Since we are presently unable to determine the relative
phases of $\fks$ and $\bpp$ as a function of $t$, the AC/GS results represent
an upper bound on $|\rsstr|$ and carry an uncertain overall sign \cite{Mu97}. 
Hence, we give only the magnitude in Table I.

With these caveats in mind, we observe that the use of a realistic, 
non-perturbative $K\bar{K}$ spectral function increases the
kaon cloud contribution to $\rsstr$ by roughly a factor of three as
compared to the ${\cal O}(g^2)$ calculation. Moreover, the non-perturbative
result approaches the scale at which the PV electron scattering experiments 
are sensitive, whereas the ${\cal O}(g^2)$ prediction is too small to be seen. 
The enhanced scale of the kaon cloud depends critically on the presence of 
the resonance structure near $t=m_\phi^2$ in {\em both} $\bpp$ and $\fks$ 
(Fig. 1). 
This structure -- obtained without relying on the {\em a priori} assumption
of pole dominance in $\GES$ or $\GEIEZ$ -- demonstrates how the resonance
contribution arises out of the $K\bar{K}$ continuum and yields a kaon
cloud result similar in magnitude to pure VMD predictions \cite{Jaf89}.

Although the kaon cloud is only one of the virtual hadronic states 
through which $s\bar{s}$ pairs may contribute to the strangeness form factors,
our result illustrates the importance of including effects to all orders
in $g$ when computing any of these hadronic contributions. In particular,
we speculate that the results of the NRQM calculation reported in Ref.
\cite{Is97}, which includes contributions from a tower
of OZI-allowed intermediate states, will be significantly modified when
effects beyond ${\cal O}(g^2)$ are included. 
Our analysis suggests that these effects, in the guise of resonant
and non-resonant meson rescattering, determine the structure and scale of 
OZI-allowed contributions.  The presently available  scattering data is not
likely to afford a model-independent, all-orders treatment of higher-mass
contributions. Nevertheless, any
model estimate of the higher-mass content of $s\bar{s}$ spectral
functions must be evaluated by comparing its kaon cloud prediction with
the results of the analysis reported here.

We gratefully acknowledge useful discussions with D. Drechsel, G. 
H{\"o}hler,  N. Isgur, and R.L. Jaffe.
This work has been supported in part by the Deutsche
For\-schungs\-ge\-mein\-schaft (SFB 201) and the German Academic Exchange 
Service (Dok\-to\-ran\-den\-sti\-pen\-dien HSP III) (HWH) and 
in part under U.S. Department of Energy contract \# 
DE-FG06-90ER40561 and a National Science 
Foundation Young Investigator Award (MJR-M).

\vspace{-0.4cm}

\begin{table}
\begin{center}~
\begin{tabular}{|cc||c|c|c|}
Moment & Scenario & $4\mks\leq t\leq4\mns$ & $4\mns\leq t$ & Total
\\\hline\hline
$\rsstr$ [fm$^2$] & ${\cal O}(g^2)$ & $-0.017$ & $-0.007$ & $-0.024 $ \\
$|\rsstr|$ [fm$^2$] & AC/GS & $0.065$ & $0.001$ & $0.066$
\end{tabular}
\vspace{0.5cm}
\caption{\label{tab1} Kaon cloud prediction for $\rsstr$ for the two
scenarios discussed in the text. Second and third columns give contributions
to the dispersion integral of Eq. (\ref{eq:rsstrdr}) from integration
regions corresponding to unphysical (second col.) and physical (third col.) 
t-channel scattering amplitudes.
In both scenarios, the unitarity bound on $\bpp$ is imposed for $t\geq 4\mns$.}
\end{center}
\end{table}

\vspace{-0.25cm}

\begin{references}

\vspace{-1.25cm}

\bibitem{Kap88} D.B. Kaplan and A. Manohar, Nucl. Phys. {\bf B310}, 527 (1988).
\bibitem{Mus94} M.J. Musolf {\em et al.}, Phys. Rep. {\bf 239}, 1 (1994)
and references therein.
\bibitem{Lei96} D.B. Leinweber, Phys. Rev. {\bf D53}, 5115 (1996); S.J. Dong,
	K.F. Liu, and A.G. Williams, hep-ph/9712483.
\bibitem{Mue97} B. Mueller {\em et al.}, 
Phys. Rev. Lett. {\bf 78}, 3824 (1997).
\bibitem{Ram97} M.J. Ramsey-Musolf and H. Ito, Phys. Rev. {\bf C55}, 3066
(1997).
\bibitem{Jaf89} R.L. Jaffe, Phys. Lett. {\bf B229}, 275 (1989); see also
H.-W. Hammer, U.-G. Mei{\ss}ner, and D. Drechsel, Phys. Lett. {\bf B367},
323 (1996).
\bibitem{Ha97} H.-W. Hammer and M.J. Ramsey-Musolf,
to appear in Phys. Lett. {\bf B}, hep-ph/9703406.
\bibitem{Mu97} M.J. Musolf, H.-W. Hammer, and D. Drechsel, Phys. Rev.
{\bf D55}, 2741 (1997).
\bibitem{Hy92} J.S. Hyslop et al., Phys. Rev. {\bf D46}, 961 (1992).
\bibitem{Ni70} H. Nielsen, J. Lyng Petersen, and E. Pietarinen,
Nucl. Phys. {\bf B22}, 525 (1970).
\bibitem{Ni72} H. Nielsen and G.C. Oades, Nucl. Phys. {\bf B41},
525 (1972).
\bibitem{Ba69} V. Barger, Phys. Rev. {\bf 179}, 1371 (1969).
\bibitem{Vi75} N.M. Queen and G. Violini, {\em Dispersion Theory in 
high-energy Physics}, (Wiley, New York, 1975), Chapter 12.
\bibitem{Br79} R.A.W. Bradford and B.R. Martin, Z. Phys. {\bf C1},
357 (1979).
\bibitem{Is97} P. Geiger and N. Isgur, Phys. Rev. {\bf D55}, 299 (1997).
\end{references}
\end{document}